%% file: new-gb.tex
\documentclass{elsart}  %
\usepackage{epsfig}
\usepackage{amssymb}
\usepackage{cite}

\begin{document}
%
\begin{frontmatter}
\title{
\hfill{\small{PITHA 98/28}}\\
\vspace{3mm}
          Universality of the gauge-ball spectrum \\
        of the four-dimensional pure U(1) gauge theory}
\author{J. Cox}
\address{Physics Department, MIT, USA} 
\author{J.~Jers{\'a}k, H. Pfeiffer}
\address{Institut f{\"u}r Theoretische Physik E, RWTH Aachen, Germany}
\author{T.~Neuhaus}
\address{Helsinki Institute of Physics, Finland}
\author{P.~W.~Stephenson}
\address{Dipartimento di Fisica and INFN, Universit\`a degli Studi di Pisa, Italy}
\author{A. Seyfried}
\address{Department of Physics, University of Wuppertal, Germany}
\date{\today}
\begin{abstract}
  We continue numerical studies of the spectrum of the pure U(1) lattice gauge
  theory in the confinement phase, initiated in \cite{CoFr97b}. Using the
  extended Wilson action $ S = -\sum_P \left [\beta \cos(\Theta_P) + \gamma
    \cos(2\Theta_P)\right ] $ we address the question of universality of the
  phase transition line in the ($\beta,\gamma$) plane between the confinement
  and the Coulomb phases. Our present results at $\gamma= -0.5$ for the
  gauge-ball spectrum are fully consistent with the previous results obtained
  at $\gamma= -0.2$. Again, two different correlation length exponents,
  $\nu_{\rm ng} = 0.35(3)$ and $\nu_{\rm g} = 0.49(7)$, are obtained in
  different channels. We also confirm the stability of the values of these
  exponents with respect to the variation of the distance from the critical
  point at which they are determined. These results further demonstrate
  universal critical behaviour of the model at least up to correlation lengths
  of 4 lattice spacings when the phase transition is approached in some
  interval at $\gamma\leq -0.2$.
\end{abstract}
\end{frontmatter}

\section{Introduction}

The properties of the strongly coupled pure U(1) lattice gauge theory (U(1)
theory) remain insufficiently understood after 20 years of its numerical
investigation. Because of the two-state signal present on finite toroidal
lattices (cubic lattices with periodic boundary conditions), it remains
unclear whether a continuum field theory can be constructed at the
confinement-Coulomb phase transition. A two-state signal is obviously
consistent with the first order, but it does not exclude the second order as
long as the latent heat decreases with increasing volume and can be
extrapolated to zero in the thermodynamic limit using a power law. Results of
recent investigations \cite{CaCr98a,CaCr98b,CoNe99} allow such an
extrapolation in some range of coupling parameters. Furthermore, when
homogeneous spherical lattices (as opposed to the surface of a 5-dimensional
hypercube) are used, the two-state signal is not observed at all
\cite{JeLa96a,JeLa96b}.

In such a situation there is little hope of answering the question of the
order beyond any doubt while no reliable analytic means are available and
lattice sizes in numerical simulations are restricted. Furthermore, the
lattice action can be considerably varied and the extended Wilson action used
in most works, and also in this paper, is only an example. There may exist
regions of the parameter space where the transition is clearly continuous.
Indeed, in a recent work \cite{Ko98b} the existence of a continuous phase
transition possibly with an essential singularity is indicated by results
using a novel formulation of the pure compact Abelian gauge theory in
continuum.

Therefore it is appropriate to ask how far the properties of the phase
transition are also consistent with the scaling behaviour associated with the
continuous transition. If for a certain action such a consistency is found in
some finite range of scales (in lattice cutoff units), two interesting
scenarios are available: First, this consistency indicates the existence of a
critical point not too distant in parameter space from the action used so
that it dominates the behaviour of the chosen system in the observed range of
scales.  Second, if the range of scales is very large, the theory might be
considered as an effective theory, with finite but large cutoff, even if no
critical point is found.

This may be considered a realistic approach to the question of whether
the U(1) theory has a physical meaning because considerable indications of a
second-order scaling behaviour for the extended Wilson action (parameterized
by two simplest plaquette couplings $\beta$ and $\gamma$) already exist. On
homogeneous spherical lattices, a volume dependence of several bulk
observables described by simple second-order finite size scaling theory was
found in a large range of lattice sizes \cite{JeLa96a,JeLa96b} at three points
of the phase transition, $\gamma = 0, -0.2, -0.5$, of the extended Wilson
action and also for the Villain action \cite{LaPe96}. This fact (and not the
absence of the two-state signal on spherical lattices) is up to now the most
significant evidence for the existence of a non-Gaussian fixed point and
universality in the U(1) theory.

The other indication of second-order scaling behaviour has been found in the
spectrum studies of the U(1) theory in the confinement phase on toroidal
lattices. Working with the extended Wilson action at $\gamma = -0.2$, the
spectrum of the gauge-balls (GB) at various distances $\tau = |\beta -
\beta_c|$ from the phase transition was determined \cite{CoFr97b}. Its
$\tau$-dependence has been found to be described in a very broad interval
(outside the narrow region where the two-state signal appears) by simple power
laws. Such a behaviour is predicted by the leading term in a second-order
scaling picture, parameterized by the correlation length exponent $\nu$. Two
different values of the exponent $\nu$ were found, one of them non-Gaussian.
This suggests the existence of an interesting fixed point in the pure U(1)
theory. Of course, another possible explanation discussed in \cite{CoFr97b} is
the occurrence of some cross-over or pre-asymptotic phenomenon. A possible
way to distinguish between these possibilities is to look for universal
properties and to repeat the calculations at a different point on the phase
boundary.

Therefore in this paper we perform a similar analysis at $\gamma = -0.5$. The
main aim is to check whether the results for $\nu$ are consistent with those
at $\gamma = -0.2$. This turns out to be the case, further supporting the
presence of a non-Gaussian fixed point, which is the most natural explanation
of such a universal scaling behaviour. The alternative explanation that the
results of \cite{CoFr97b} are influenced by a crossover related to a possible
tricritical point in the vicinity is no longer probable.

We pay particular attention to the stability of the values of $\nu$ with
respect to the variation of the $\beta$-interval in which they are determined
by power-law fits to the measured masses. We demonstrate, also for data
obtained previously \cite{CoFr97b} at $\gamma = -0.2$, that, within the
lattice sizes and correlation lengths we could achieve, there is no
significant indication that $\nu$ approaches the value 1/4, expected for a
weak first order phase transition.

The new results are presented as follows: In section~2 we briefly describe the
model and summarize our earlier results on the GB spectrum at $\gamma=-0.2$.
In section~3 we present results for the scaling behaviour of the GB masses at
$\gamma=-0.5$. In section~4 the stability of the results at both $\gamma$
values is discussed. In section~5 we compare the results at both $\gamma$
values and draw our conclusions.

\pagebreak

\section{The U(1) theory and earlier results for GB masses at $\gamma = -0.2$}

\subsection{Action and phase diagram}

We consider pure U(1) lattice gauge theory with an extended Wilson action
\begin{equation}\label{ACTION}
         S = -\sum_P
              \left [\beta \cos(\Theta_P) + \gamma
                \cos(2\Theta_P)\right ].
\end{equation}
Here, $\Theta_P \in [0,2\pi)$ is the plaquette angle, i.e. the
argument of the product of U(1) link variables around a plaquette $P$,
and $\beta$ and $\gamma$ are the single and the double charge
representation couplings, respectively. Taking $\Theta_P =
a^2gF_{\mu\nu}$, where $a$ is the lattice spacing and $\beta +
4\gamma = 1/g^2$, one obtains for weak coupling $g$ the usual
continuum action $S =\frac{1}{4} \int d^4xF_{\mu\nu}^2$.

This lattice gauge theory has a line of phase transitions between the strong
coupling confinement phase and the weak coupling Coulomb phase.  Its position
for the Wilson action ($\gamma = 0$) is $\beta_c \simeq 1.011$ \cite{EvJe85},
whereas for $\gamma = -0.5$ it is given below in (\ref{BETA_C}). For $\gamma
\ge +0.2$ the transition is most probably of $1^{\rm st}$ order, weakening
with decreasing $\gamma$ \cite{Bh82,EvJe85}.

The recent studies on spherical lattices suggest that the order changes at
$\gamma = \gamma_0 \simeq 0$, and is of $2^{\rm nd}$ order for $\gamma \le
\gamma_0$ \cite{JeLa96a,JeLa96b}. With decreasing $\gamma$ the $2^{\rm nd}$
order transition further weakens in the sense that the specific heat peak
decreases for fixed lattice size and the autocorrelation time increases
\cite{JeLa96b}. The scaling behaviour of bulk quantities is universal at least
in the range $-0.5 \le \gamma \le 0$.

On toroidal lattices the disturbing two-state signal weakens with
decreasing $\gamma$, but is present at least until $\gamma = -0.5$
\cite{EvJe85}.  For even smaller $\gamma$ the large autocorrelation
time makes simulations prohibitively expensive. Thus on the toroidal
lattices the two-state signal on the critical line cannot be avoided.

\subsection{Summary of the earlier results at $\gamma=-0.2$}

Now we briefly summarize those results obtained for $\gamma = -0.2$ in
\cite{CoFr97b} which are relevant for this work. In the vicinity of the
critical point 
\begin{equation}
  \beta_c=1.1609(2)
\end{equation}
in the confinement phase two groups of GB masses have been found with
distinctly different scaling behaviour when the phase transition is
approached, $\tau \to 0$. Most of the GB masses scale proportionally to
$\tau^{\nu_{\rm ng}}$, the value of the correlation length exponent $\nu_{\rm
  ng}$ being non-Gaussian,
\begin{equation}
\nu_{\rm ng} =  0.35(3).
\label{NUNG}
\end{equation}
This agrees with the exponent $\nu = 0.365(8)$ found in FSS studies on
spherical lattices \cite{JeLa96a,JeLa96b,LaPe96}.

The mass of the $J^{PC} = 0^{++}$ GB scales as 
$\tau^{\nu_{\rm g}}$, with
\begin{equation}
         \nu_{\rm g} =  0.49(7).
\label{NUG}
\end{equation}
This correlation length exponent agrees with the Gaussian value
$\nu = 1/2$. We also give the result for the ratio of
both exponents,
\begin{equation}
         \frac{\nu_{\rm ng}}{\nu_{\rm g}} =  0.71(8).
\label{NU_RATIO}
\end{equation}

We note that an observation of two different critical exponents at one value
of $\gamma$ leaves open the possibility that one of these exponent values is
caused by crossover phenomena associated with the possible presence of a
tricritical point in the vicinity. Two such scenarios have been discussed in
\cite{CoFr97b}. Their confirmation or rejection requires repeating the
calculations either at the same $\gamma$ and substantially larger lattices,
which is currently not possible, or at another, distant point in the parameter
space.

\section{Analysis at $\gamma = -0.5$}

\subsection{Statistics}

We have used a vectorized three-hit Metropolis algorithm with an acceptance of
about 50\% at each hit. Due to the larger autocorrelation at $\gamma=-0.5$
compared with $\gamma=-0.2$ the measurements have been done after 50
Metropolis sweeps each.

\input{tab/statistics}

Table~\ref{tab:stat} shows the lattice sizes $L_s^3\cdot L_t$ with
$L_t=2\,L_s$, the $\beta$ values of the simulations, and the number of
measurements in multiples of $1000$. The phase transition turns out to be
located at $\beta_c=1.4223(2)$. 

The dagger \dag\ in table~\ref{tab:stat} denotes the simulations in which the
two state signal is observed in the evolution of the plaquette energy. These
runs have been excluded from the analysis. 

The measurements of the GB masses use the techniques of smearing and
diagonalization of the correlation matrices described for pure QCD in
\cite{Te86,MiTe89}. The measurement of the effective energies $\epsilon_j(t)$
in the various GB channels and the way of estimating their statistical errors
is explained in section~4 of~\cite{CoFr97b}. 

\subsection{Gauge-ball channels}

The various gauge-ball channels result from the irreducible representations of
the proper cubic group labeled $R=A_1,A_2,E,T_1,T_2$ (with dimensions
$1,1,2,3,3$, respectively). Inversion then introduces a parity quantum number
$P=\pm 1$ and, since the representations are complex, one can also assign a
charge parity $C=\pm 1$. The GB states are completely classified by $R^{PC}$
\cite{BeBi83}.

We measured the correlation functions of all the $20$ channels $R^{PC}$ with
momentum $p=0$ and some of them with the smallest lattice momentum $p=1$ (in
units of $2\pi/L_s$). For the construction of operators with a certain
symmetry see section~4 of \cite{CoFr97b} as well as \cite{BeBi83,Mi90}.

The effective energies $\epsilon_j(t)$ in each GB channel $j=R^{PC}$ have been
determined for as many $t\leq 6$ as possible. The GB energies $E_j$ can in
principle be obtained from a fit to a plateau in the values of $\epsilon_j(t)$
which is expected for large $t$. Since the errors are sometimes very large or
the effective energies could not be determined at all, the reliability of the
determination of the energies $E_j$ strongly depends on the GB channel. We
classify this reliability in a similar way as done in \cite{CoFr97b}: 

\begin{description}
\item [\bf $+++$ states.] 
  At most $\beta$ values a plateau in $\epsilon_j(t)$ is found
  and can be fitted at $t\geq 2$ or $t\geq 3$.
\item [\bf $++$ states.]
  At most $\beta$ values a plateau in $\epsilon_j(t)$ is found
  only if $t=1$ is included.
\item [\bf $+$ states.]
  Plateaus in $\epsilon_j(t)$ can be found only at some $\beta$
  values and only if $t=1$ is included.
\item [\bf $?$ states.]
  No plateaus in $\epsilon_j(t)$ are found: we cannot determine a GB energy
  at all.  
\end{description}

\input{tab/symmetries}

Table~\ref{tab:channels} shows the quality of the GB channels at
$\gamma=-0.5$. It is similar to the situation at $\gamma=-0.2$ \cite{CoFr97b}
with some differences in the $+$ and $?$ states. In contrast to
\cite{CoFr97b} we have no $\beta$ point close to the phase transition with a
particularly high statistics. This prevents us from verifying small deviations
from the one-particle dispersion relation for states denoted in \cite{CoFr97b}
by an asterisk.

\begin{figure}[tbp]
  \centerline{
    \psfig{file=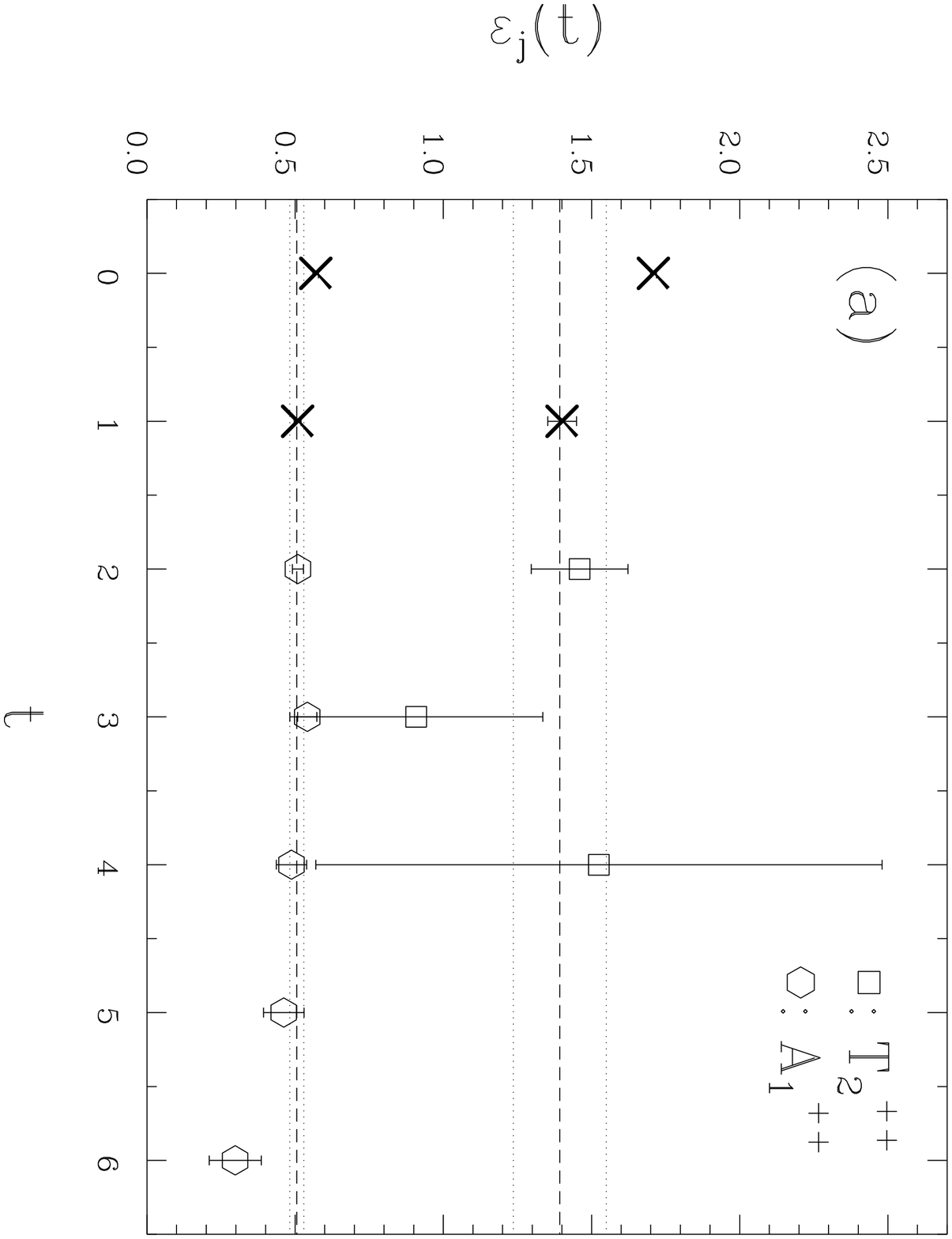,angle=90,width=7.2cm}
    \psfig{file=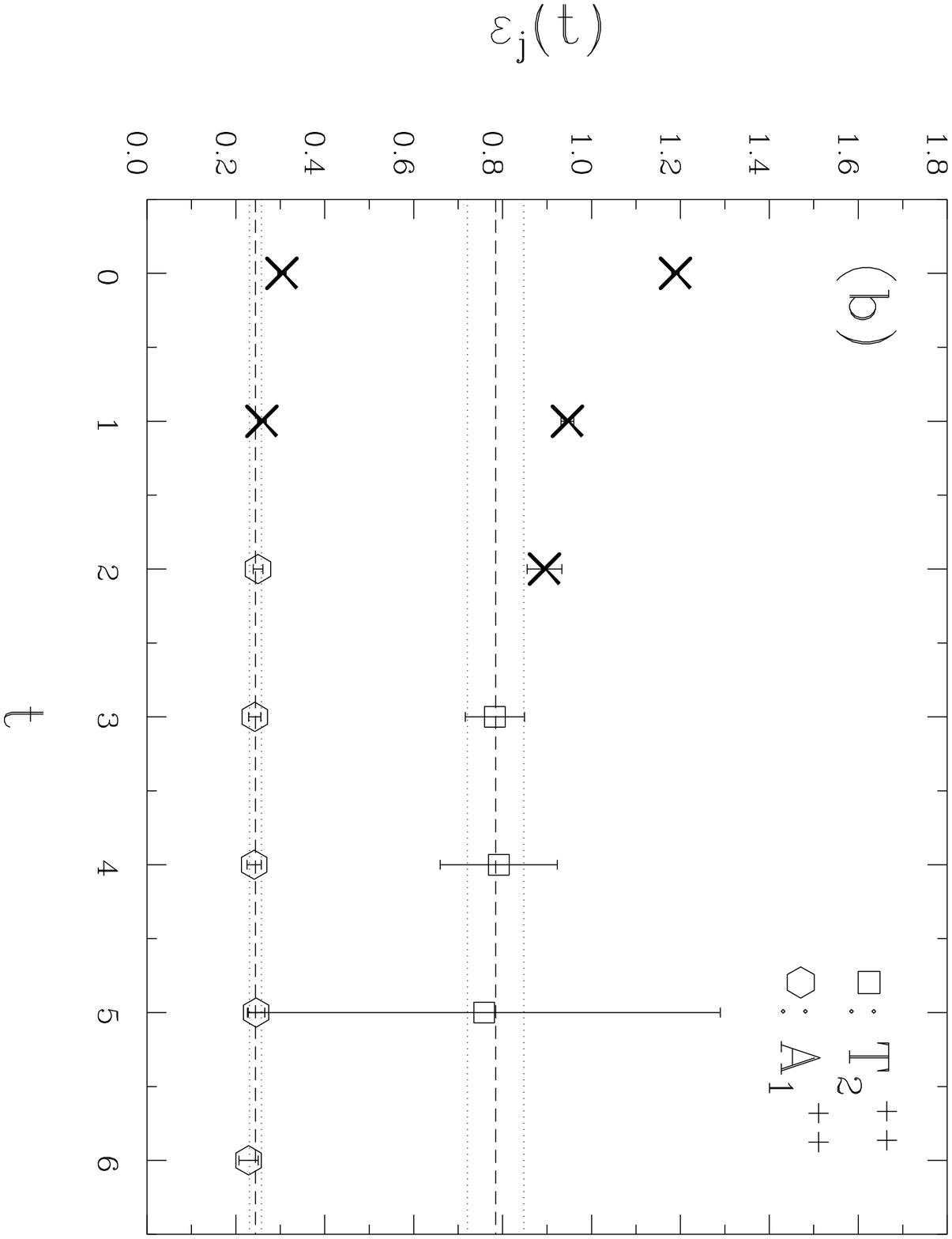,angle=90,width=7.2cm}
    }
  \caption{Effective energies from $A_1^{++}(0)$ and $T_2^{++}(0)$ correlation
    functions (a)~at $\beta=1.414$ on the $16^3 32$ lattice and (b)~at
    $\beta=1.42$ on the $24^3 48$ lattice. The plateaus have been fitted
    omitting the points at $t=0,1$ at $\beta=1.414$, omitting $t=0,1$ for
    $A_1^{++}$ and $t=0,1,2$ for $T_2^{++}$ at $\beta=1.42$. The resulting GB
    energies $E_j$ and their errors are indicated by horizontal lines.}  
  \label{fig:effmass}
\end{figure}

The effective energies $\epsilon_j(t)$ of $A_1^{++}$ ($+++$ quality) and
$T_2^{++}$ ($++$ quality) with $p=0$ are shown in fig.~\ref{fig:effmass} (a)~at
$\beta=1.414$ on the $16^3 32$ lattice and (b)~at $\beta=1.42$ (close to
$\beta_c$) on the $24^3 48$ lattice. The horizontal lines are the resulting GB
energies $E_j$ with errors.

The GB masses $m_j$ have been calculated from the GB energies $E_j$ using the
lattice dispersion relation
\begin{equation}
  2\,(\cosh E_j-1)=m_j^2+2\,\sum_{\mu=1}^3(1-\cos p_\mu).
\end{equation}
If the masses for a certain $R^{PC}$ are consistent for both $p=0$ and $p=1$,
we assumed equality and used the same parameter in the fits. In this case both
momenta are listed in the same line in table~\ref{tab:channels} and are
denoted by the same symbol in figs.~\ref{fig:nu}
and~\ref{fig:spectrum}. 

\subsection{Two mass scales}

Having determined the GB masses $m_j$ at various $\beta$, their scaling
behaviour with $\beta$ is analyzed. First, the power law
\begin{equation}
  m_j=c_j{(\beta_c^{(j)}-\beta)}^{\nu_j}
\end{equation}
is fitted to the individual GB masses measured in each GB channel
$j=R^{PC}$. The results for the critical values $\beta_c^{(j)}$ agree well
within the error bars, so we assume equality $\beta_c^{(j)}=\beta_c$ and use
the power law 
\begin{equation}
\label{eq:power}
  m_j=c_j{(\beta_c-\beta)}^{\nu_j}
\end{equation}
with one $\beta_c$ for all GB channels.

\begin{figure}[tbp]
  \centerline{
    \psfig{file=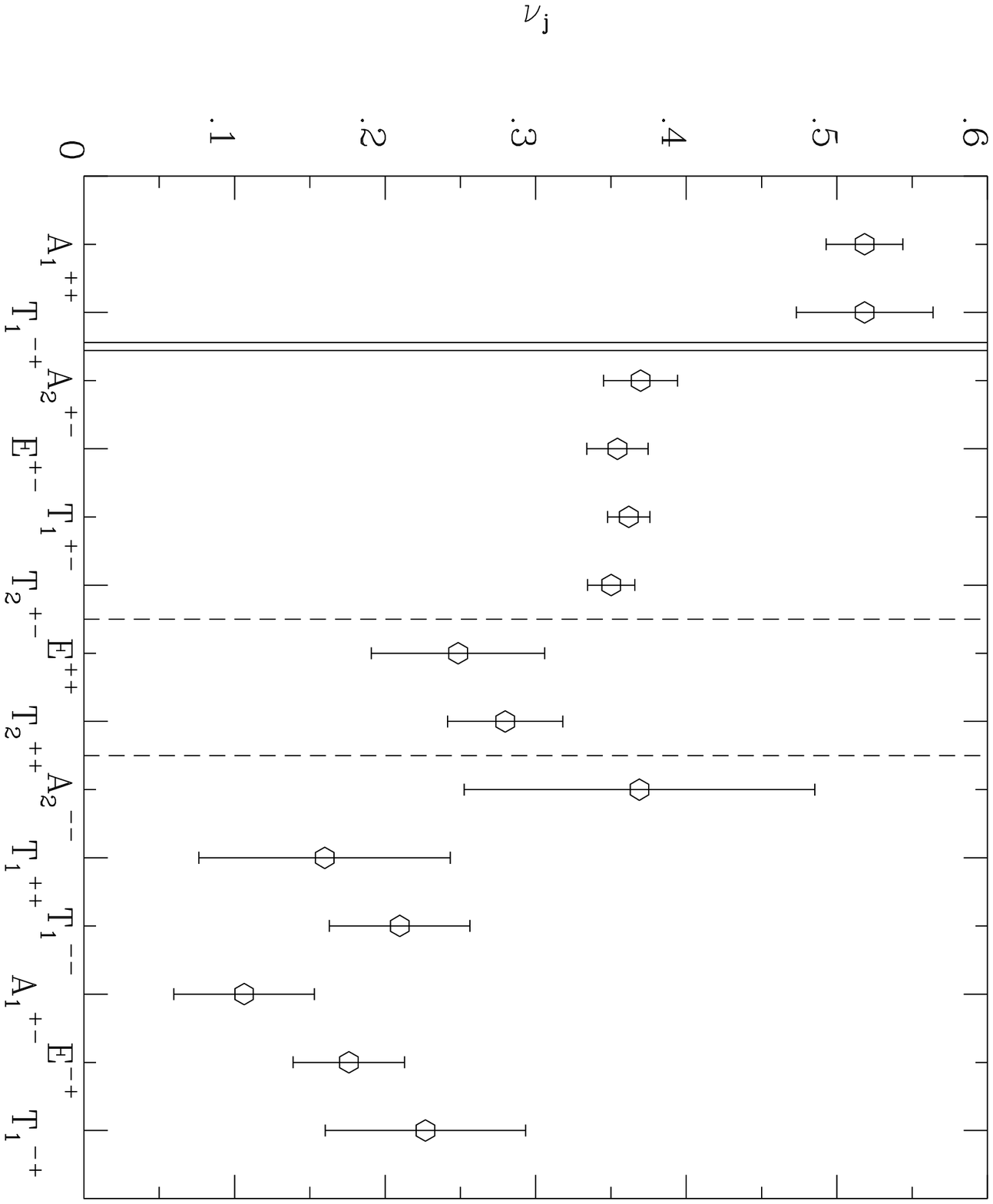,angle=90,width=10cm,bbllx=21,bblly=180,bburx=588,bbury=828}
    }
  \caption{Values of $\nu_j$ obtained separately in each channel. The
    double vertical line separates two groups with distinctly different
    $\nu_j$. The dashed vertical lines separate channels with different
    quality within the group with the non-Gaussian exponent with decreasing
    quality from left to right.}
  \label{fig:nu}
\end{figure}

The resulting $\nu_j$ from the fits~(\ref{eq:power}) are shown in
fig.~\ref{fig:nu}. In these fits we restricted $\beta$ to the values having 
$m_j(\beta)\leq 1.0$ in lattice units. So the scaling behaviour could be
determined for all GB channels with $+++$ and $++$, but only for some of the
$+$ channels.

Fig.~\ref{fig:nu} shows different critical exponents $\nu_j$ for the scaling
of GB masses in two groups of GB channels separated by the double vertical
line in the figure. These two groups consist of the same GB channels as found
at $\gamma=-0.2$ \cite{CoFr97b}. The two critical exponents are very similar
to the results at $\gamma=-0.2$, see~(\ref{NUNG}),(\ref{NUG}), and are
therefore called $\nu_{\rm g}$ and $\nu_{\rm ng}$ respectively. The third
column of table~\ref{tab:channels} shows the association of GB channels to one
of the two groups with $\nu_{\rm g}$ or $\nu_{\rm ng}$.

\begin{figure}[tbp]
  \centerline{
    \psfig{file=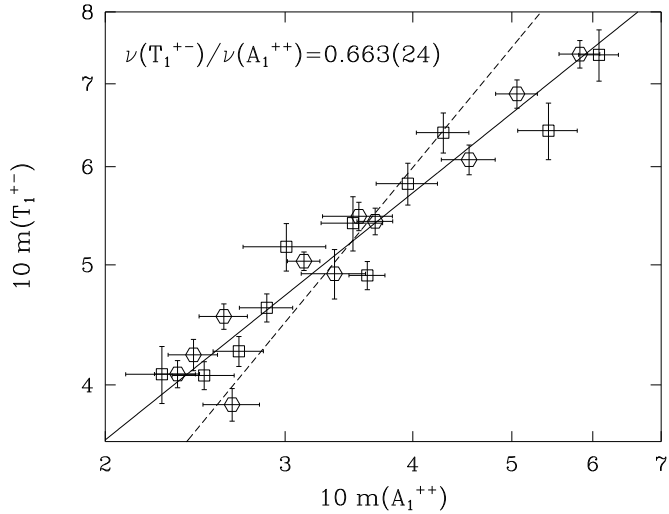,angle=90,width=10cm,bbllx=21,bblly=15,bburx=588,bbury=828}
    }
  \caption{A log-log plot of the masses of the $A_1^{++}$
    and $T_1^{+-}$ gauge-balls in the interval $1.412\leq \beta\leq
    1.4205$. The results from the channels with $p=0,1$ are denoted by
    circles and squares, respectively. The solid line is the result of
    a straight line fit considering errors in both directions. The
    dashed line with a gradient of one, i.e. assuming equal
    exponents, is shown for comparison. }
  \label{fig:loglog}
\end{figure}

To demonstrate the difference between the exponents $\nu_{\rm g}$ and 
$\nu_{\rm ng}$ in a way independent of the numerical value $\beta_c$, 
fig.~\ref{fig:loglog} shows the masses of $A_1^{++}$ and $T_1^{+-}$ in a
doubly logarithmic plot. These two channels are the highest quality
representatives of the $\nu_{\rm g}$ and $\nu_{\rm ng}$ group, respectively. 

The slope of the solid line in fig.~\ref{fig:loglog} is the ratio 
\begin{equation}
\label{NU_RATIO5}
\frac{\nu_{\rm ng}}{\nu_{\rm g}}=0.663(24).
\end{equation}
The dashed line with slope one (if the exponents were equal) is shown for
comparison. The result~(\ref{NU_RATIO5}) is compatible with the 
value~(\ref{NU_RATIO}) obtained at $\gamma=-0.2$.

\begin{figure}[tbp]
  \centerline{
    \psfig{file=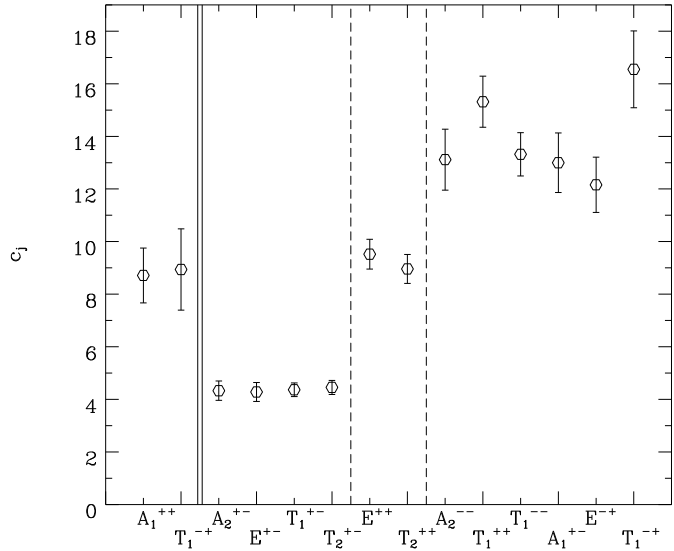,angle=90,width=10cm,bbllx=21,bblly=180,bburx=588,bbury=828}
    }
  \caption{The amplitudes $c_j$ of the fits~\protect{(\ref{eq:power})}
    under the assumption that there are just two critical exponents. The
    vertical lines have the same meaning as in \protect{fig.~\ref{fig:nu}}.} 
  \label{fig:spectrum}
\end{figure}

Assuming there are just two critical exponents $\nu_{\rm g}$ and 
$\nu_{\rm ng}$ we use the power law fit~(\ref{eq:power}) again with only two
different exponents $\nu_j$. Fig.~\ref{fig:spectrum} shows the results obtained
for the amplitudes $c_j$. We find the same amplitude for all $\nu_{\rm g}$ GB
masses. The amplitudes of the $\nu_{\rm ng}$ GB masses seem to be integer
multiples of another amplitude value. This situation is again
very similar to the results at $\gamma=-0.2$ \cite{CoFr97b}.

The ratios of the amplitudes $c_j$ in each of the two groups of GB states have
been determined using the amplitudes of the highest quality channels
$A_1^{++}$ and $T_1^{+-}$ as a reference. These amplitude ratios determine the
mass ratios in a possible continuum limit and are shown in the fifth column of
table~\ref{tab:channels}. The results at $\gamma=-0.2$ from \cite{CoFr97b} are
given for comparison. The amplitude ratios turn out to be compatible 
within error bars for all GB channels for which the amplitudes have been
determined at both $\gamma$ values.

To present results for the values of $\beta_c$, $\nu_{\rm g}$, and 
$\nu_{\rm ng}$ with reliable error estimates, we repeat the power law
fit~(\ref{eq:power}) with two exponents and fit the $+++$ gauge-balls
only. The results are 
\begin{equation}
\label{NUG5}
  \nu_{\rm g}=0.51(3),
\end{equation}
\begin{equation}
\label{NUNG5}
  \nu_{\rm ng}=0.35(2),
\end{equation}
and
\begin{equation}
\label{BETA_C}
  \beta_c=1.4223(2).
\end{equation}
The ratio $\nu_{\rm ng}/\nu_{\rm g}$ is consistent with~(\ref{NU_RATIO5}).

\begin{figure}[tbp]
  \centerline{
    \psfig{file=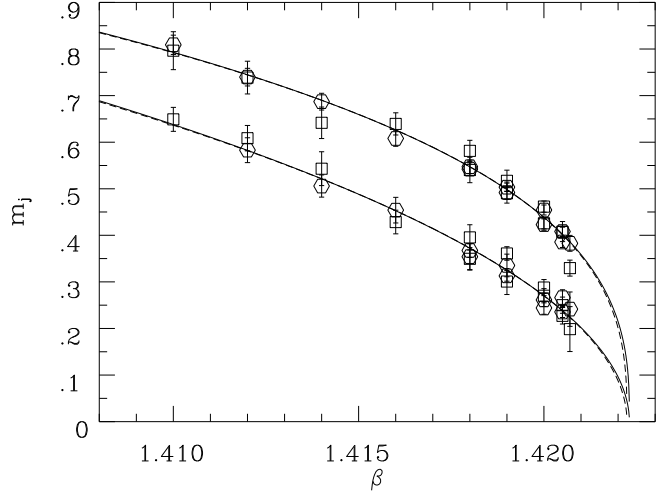,angle=90,width=10cm,bbllx=21,bblly=15,bburx=588,bbury=828}
    }
  \caption{Masses of the $A_1^{++}$ (lower points) and $T_1^{+-}$
    gauge-balls versus $\beta$ for both momenta ($p=0,1$ being denoted by
    circles and squares, respectively). The dashed curves are fits to the data
    shown. The full curves represent the mass scales obtained in a fit with all 
    $+++$ GB states.}
  \label{fig:2scale}
\end{figure}

Fig.~\ref{fig:2scale} shows the scaling of the masses of the GB states
$A_1^{++}$ and $T_1^{+-}$ with $\beta$. The dashed curves are power law fits
to the data shown while the solid curves represent the results from the fit to
all $+++$ GB states which lead to~(\ref{NUG5})$\ldots$(\ref{BETA_C}).

\pagebreak

\section{Stability of $\nu$ and comparison with essential singularity}

\subsection{Stability of $\nu$}

\begin{figure}[tbp]
  \centerline{
    \psfig{file=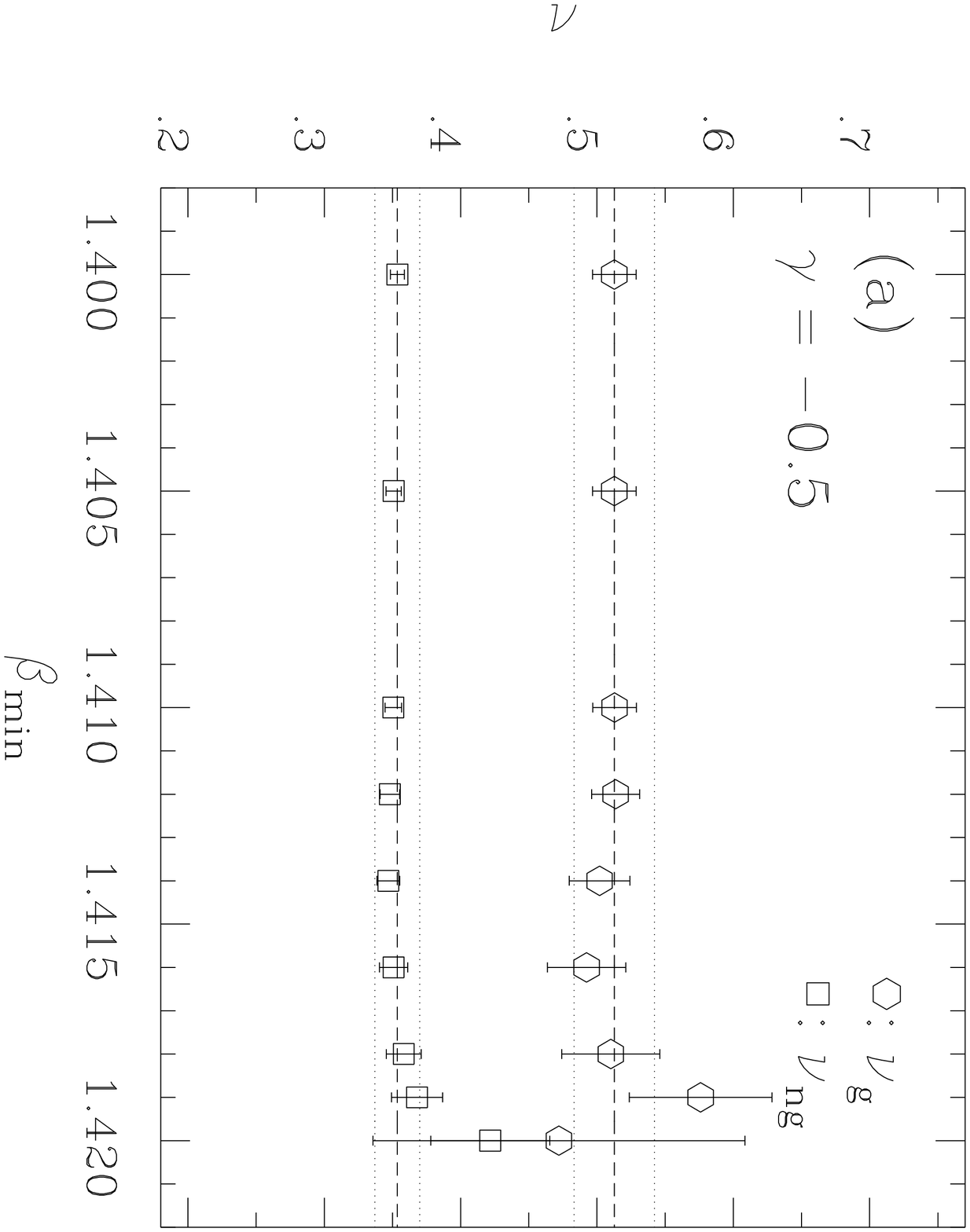,angle=90,width=7.2cm}
    \psfig{file=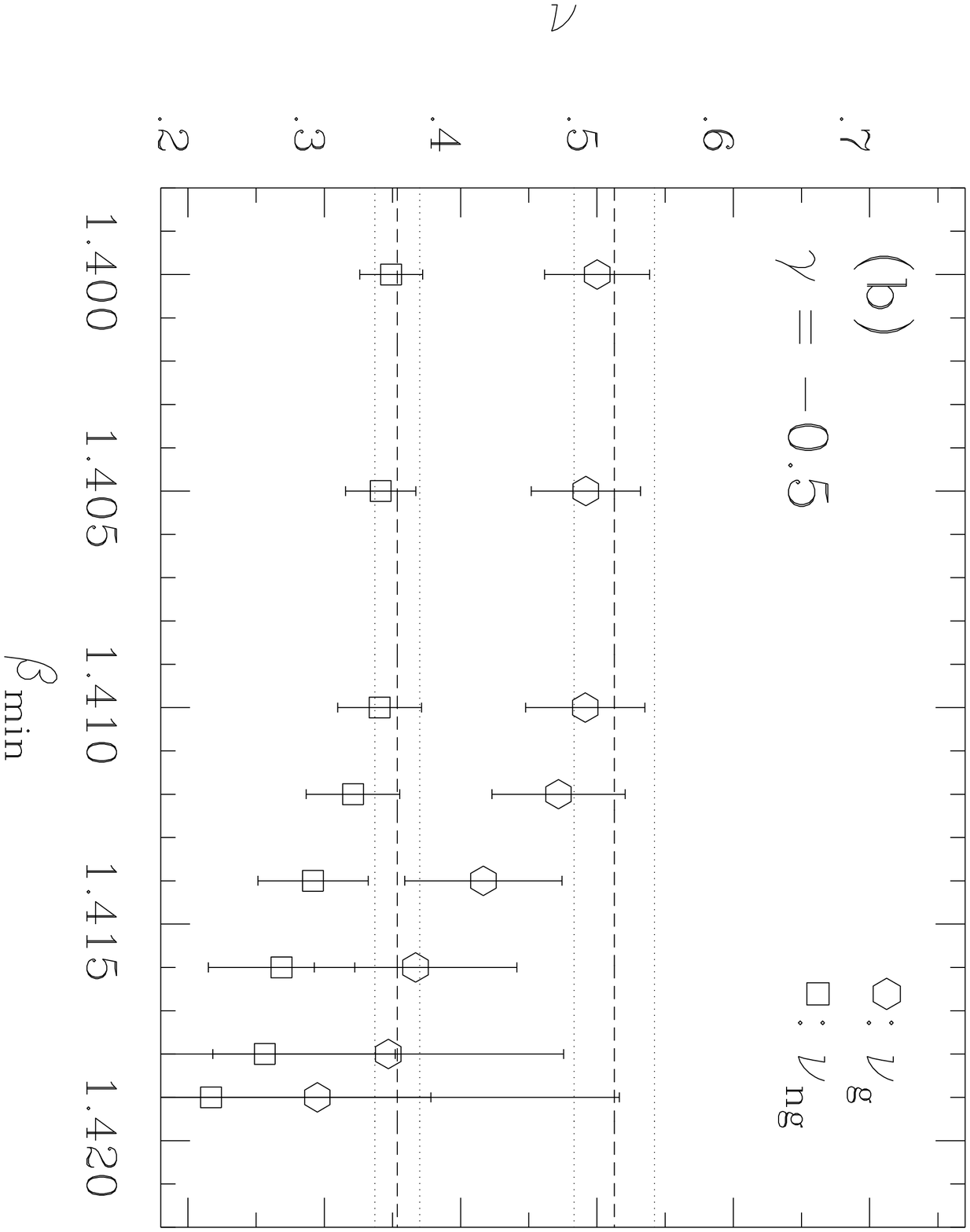,angle=90,width=7.2cm}
    }
  \caption{Values of $\nu_{\rm g}$ (circles) and $\nu_{\rm ng}$ (squares) at
    $\gamma=-0.5$ obtained in a fit with $\beta_{\rm min}\leq\beta\leq\beta_c$
    (a) with $\beta_c=1.4223$ fixed and (b) with free $\beta_c$. The
    horizontal lines indicate the values~\protect{(\ref{NUG5})}
    and~\protect{(\ref{NUNG5})}.} 
  \label{fig:stab_5}
\end{figure}

In order to test the reliability of the numerical determination of the
exponents $\nu_{\rm g}$ and $\nu_{\rm ng}$ we repeat the power law
fits~(\ref{eq:power}) and restrict the $\beta$ values to the range
near the phase transition $\beta_{\rm min}\leq\beta\leq\beta_c$. The resulting
exponents are then plotted against $\beta_{\rm min}$ to detect any possible
non-asymptotic behaviour.

Fig.~\ref{fig:stab_5}, part~(a) shows the exponents $\nu_{\rm g}$ and
$\nu_{\rm ng}$ against $\beta_{\rm min}$ obtained from the power law
fit~(\ref{eq:power}) using the $+++$ GB masses with $\beta_c=1.4223$ fixed. In
the fit with free $\beta_c$ in part~(b) of fig.~\ref{fig:stab_5} we use only
the masses in the two highest quality GB channels $A_1^{++}$ and $T_1^{+-}$ to
avoid any influence caused by the fact that the masses of some GB states
could be determined only at a few $\beta$ values near the phase transition. 

The critical exponents $\nu_{\rm g}$, $\nu_{\rm ng}$, and even their
statistical errors are very stable with respect to $\beta_{\rm min}$ when
$\beta_c$ is fixed (fig.~\ref{fig:stab_5}~(a)). With free $\beta_c$
(fig.~\ref{fig:stab_5}~(b)) the exponents are smaller if the fit is
restricted to the $\beta$ values near the phase transition, but their
statistical errors grow by an amount comparable to this effect.

\begin{figure}[tbp]
  \centerline{
    \psfig{file=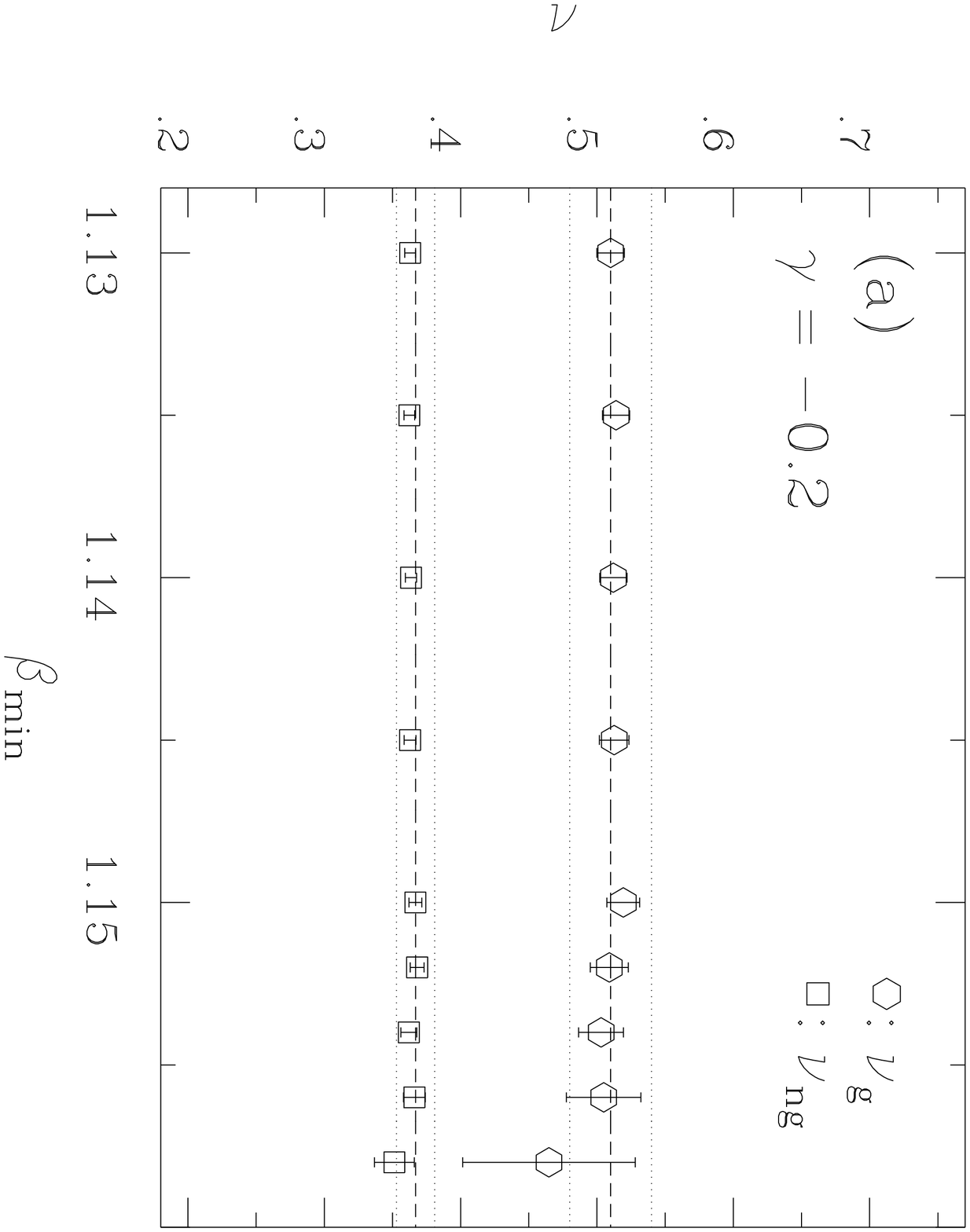,angle=90,width=7.2cm}
    \psfig{file=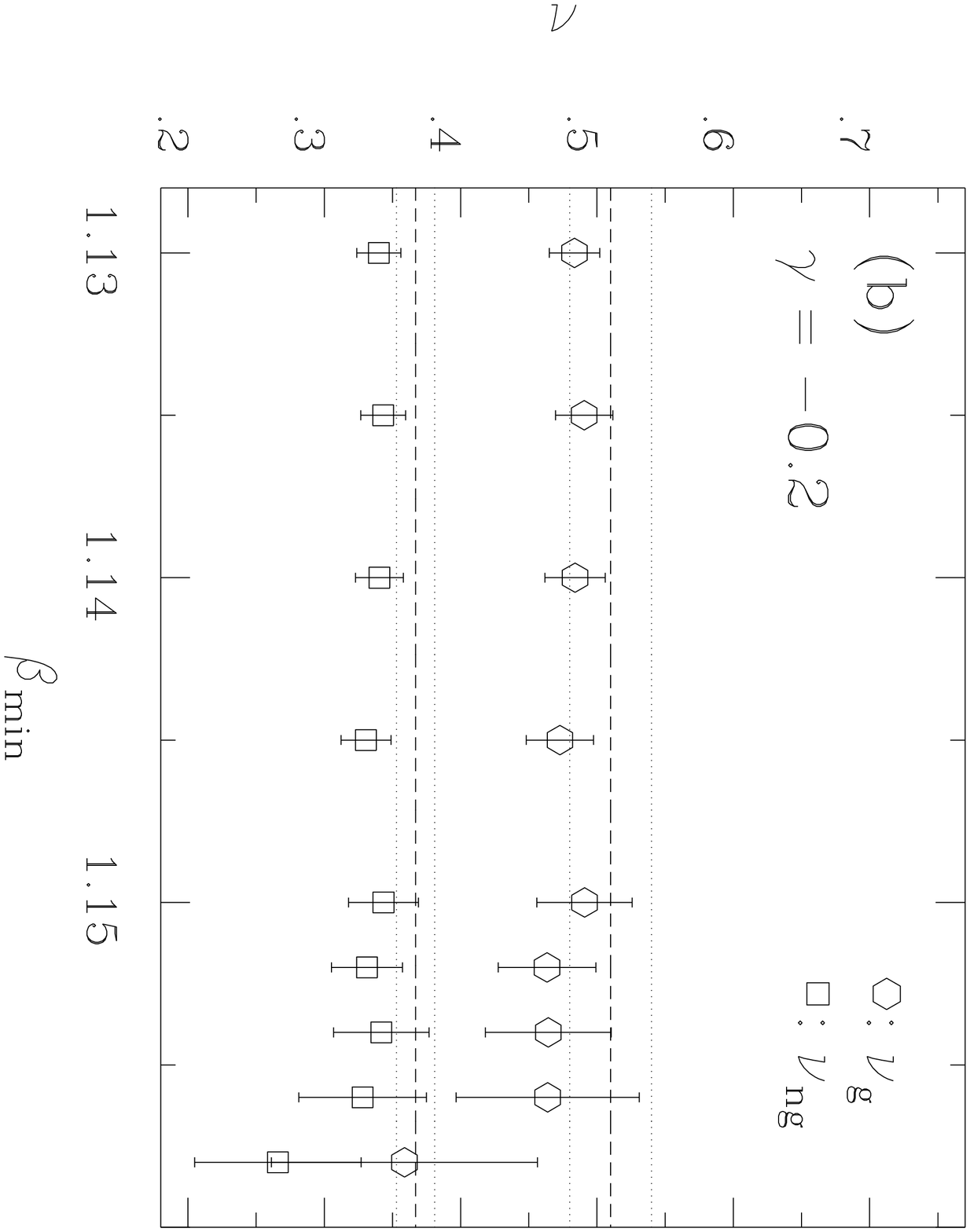,angle=90,width=7.2cm}
    }
  \caption{Values of $\nu_{\rm g}$ (circles) and $\nu_{\rm ng}$ (squares) at
    $\gamma=-0.2$ obtained in a fit with $\beta_{\rm min}\leq\beta\leq\beta_c$
    (a) with $\beta_c=1.1609$ fixed and (b) with free $\beta_c$. The
    horizontal lines indicate the values~\protect{(\ref{NUG})}
    and~\protect{(\ref{NUNG})}.}  
  \label{fig:stab_2}
\end{figure}

For comparison we also present the same analysis at $\gamma=-0.2$ using our
data from \cite{CoFr97b}. Fig.~\ref{fig:stab_2}~(a) and~(b) show these
results with $\beta_c=1.1609$ fixed and with free $\beta_c$ using the same GB
channels as in fig.~\ref{fig:stab_5}.  

The resulting critical exponents obtained with $\beta_c$ fixed turn out to be
very stable. They are quite stable even with free $\beta_c$.

Currently we are unable to decide whether the decrease of the exponents at
$\gamma=-0.5$ with free $\beta_c$ near the transition is a statistically
significant effect. The data at $\gamma=-0.2$ have in general larger
statistics than those at $\gamma=-0.5$ and do not show such a decrease
(fig.~\ref{fig:stab_2}~(b)).

If the fits are made using $g-g_c$ instead of $\beta-\beta_c$
in~(\ref{eq:power}), we find fully consistent values for the critical
exponents $\nu_{\rm g}$ and $\nu_{\rm ng}$.

\subsection{Comparison with essential singularity}

\begin{figure}[tbp]
  \centerline{
    \psfig{file=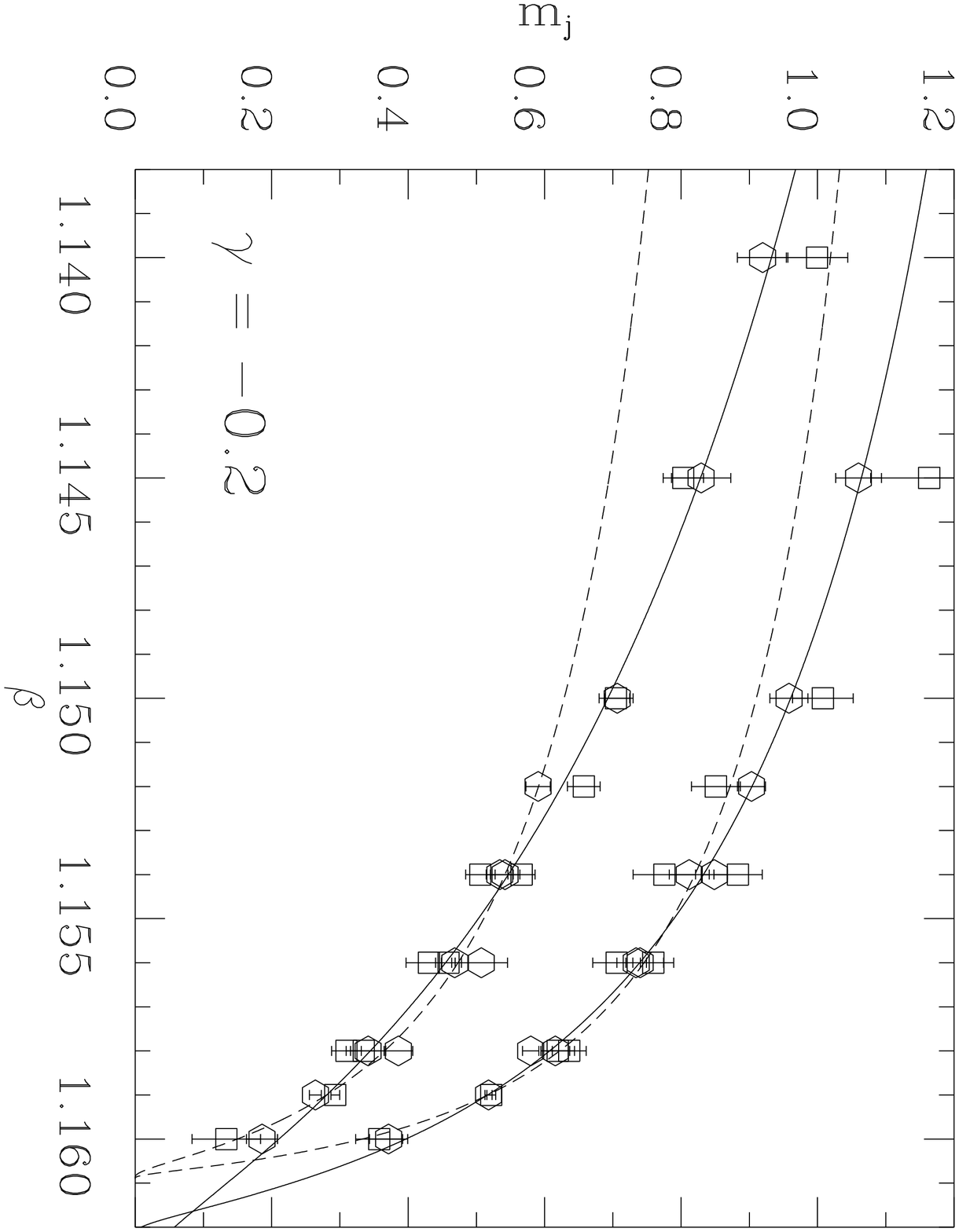,angle=90,width=7.2cm}
    \psfig{file=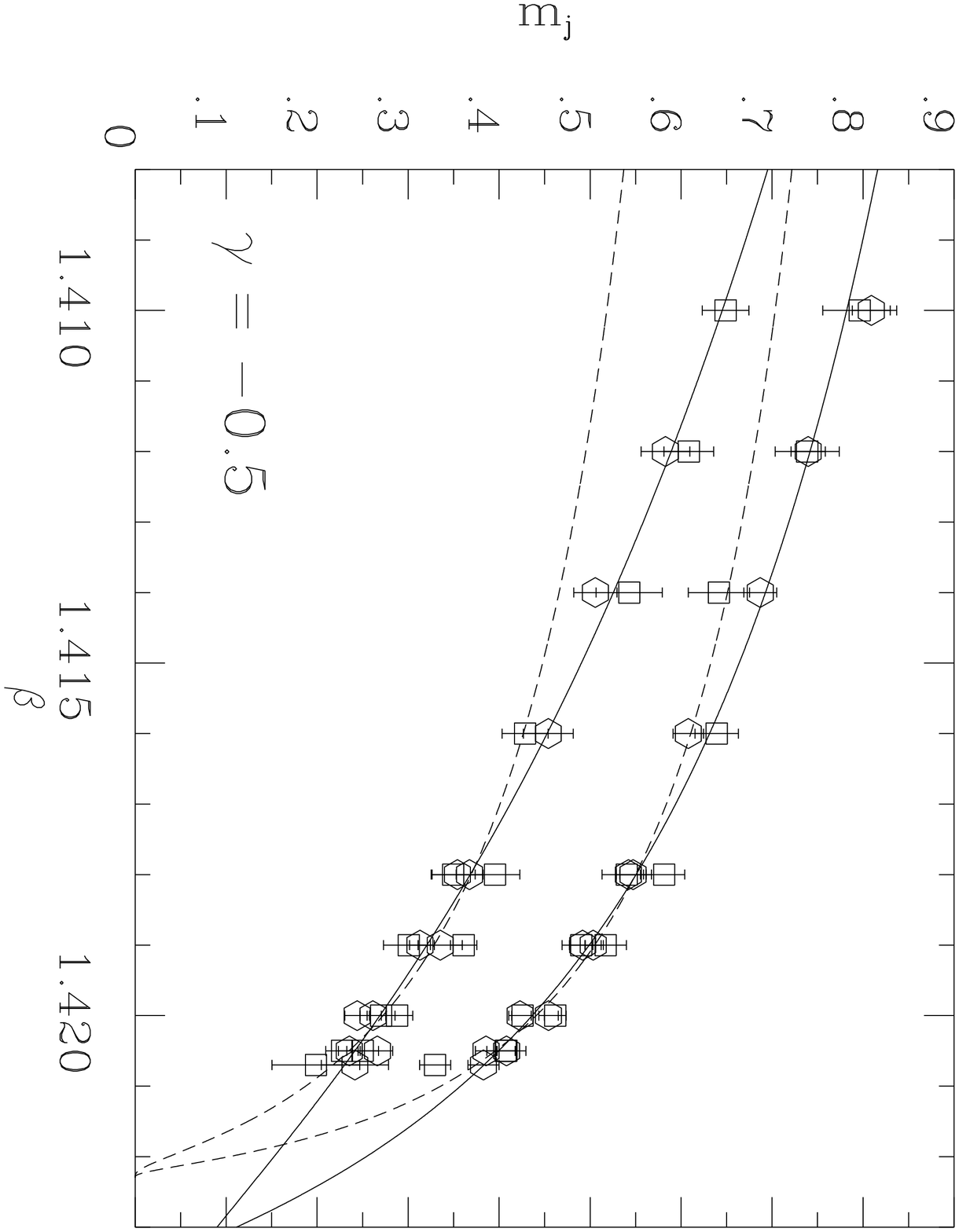,angle=90,width=7.2cm}
    }
  \caption{Masses of the $A_1^{++}$ (lower points) and $T_1^{+-}$ gauge-balls
    (upper points) at $\gamma=-0.2$ (left plot) and $\gamma=-0.5$ (right
    plot). The momenta $p=0,1$ are denoted by circles and squares,
    respectively. The fits with the essential singularity
    law~\protect{(\ref{ESS})} with free $\beta_c$ using all data give the
    solid curves. The dashed curves are produced with fixed $\beta_c$ and
    restricting $1.152\leq\beta\leq\beta_c$ at $\gamma=-0.2$ resp.\
    $1.416\leq\beta\leq\beta_c$ at $\gamma=-0.5$.}
\label{fig:ess}
\end{figure}

Motivated by \cite{Ko98b} we have also explored the possible existence of an
essential singularity and tried to describe the scaling of the masses of
$j=A_1^{++}$ and $j=T_1^{+-}$ with the essential singularity law
\begin{equation}
\label{ESS}
  m_j(\beta)=c_j\,\exp\Bigl(-\frac{\eta_j}{\sqrt{\beta_c-\beta}}\Bigr)
\end{equation}
with constants $\eta_j$.

We studied several fits. Including all data and leaving all parameters
of~(\ref{ESS}) free, the resulting $\beta_c$ is far off the anticipated value
(solid curves in fig.~\ref{fig:ess}). Fixing $\beta_c=1.4223$
(resp.~$\beta_c=1.1609$ at $\gamma=-0.2$) and restricting~(\ref{ESS}) to the
data near the phase transition $1.416\leq\beta\leq\beta_c$
(resp.~$1.152\leq\beta\leq\beta_c$ at $\gamma=-0.2$) we obtain a satisfactory
description of the masses (dashed curves). But in this case the results
$\eta_j$ are not very stable with respect to the choice of $\beta_c$, e.g.\
varying $\beta_c=1.4223\pm0.0002$ results in a variation of 
$\eta_{\rm g}=0.05\pm0.02$. 

It is thus possible that for data in the critical region the results are
consistent with an essential singularity.  However, establishing the existence
of such a singularity would need much larger correlation lengths.

\section{Comparison of the results at $\gamma = -0.2$ and $\gamma = -0.5$ and 
conclusions}

We have compared the scaling of the gauge-ball masses when the phase
transition is approached at $\gamma=-0.5$ from the confinement phase with the
analogous results obtained at $\gamma=-0.2$ \cite{CoFr97b}. Both the critical
exponents and the amplitudes ratios of the GB masses which are expected to
show a universal behaviour at a second order phase transition, have fully
compatible values at both $\gamma$.

So it is most probable that the results obtained for the scaling of GB masses
reflect properties of the line of deconfinement phase transitions in the
parameter space ($\beta,\gamma$) of the extended Wilson action in a region of
values $\gamma\leq-0.2$. The most natural explanation of this universality
would be the presence of a continuous phase transition in the coupling
parameter space not far from, though not necessarily at the points we have
investigated. Its localization may finally require the introduction of
additional coupling parameters outside the $(\beta,\gamma)$ plane of the
extended Wilson action.

This is to be compared with other attempts to explain the occurrence of
two distinct critical exponents. Certain scenarios (called $T_{\rm g}$ and
$T_{\rm ng}$ in \cite{CoFr97b}) in which the cross-over near a tricritical
point is responsible for the different exponents observed in non-vanishing
distance from the critical line, are most probably ruled out as such an effect
should be $\gamma$-dependent.

The scenario suggested in \cite{CaCr98b}, in which $\nu_{\rm ng}$ is some
effective value of $\nu$ which decreases to $1/4$ as the phase transition is
approached, is not supported (but also not excluded) by our stability
analysis. In any case $\nu_{\rm g}\simeq 0.5$ is far away from the value
$1/4=1/d$ expected for a first order transition.

Of course, future work on larger lattices may disclose new facts. The question
whether there is a certain range of parameters in which the phase transition
is of second order is still open. However, our results obtained for the
gauge-ball masses as small as $0.2$ in lattice units are fully consistent with
this possibility.

\begin{ack}
  The computations have been performed on the Cray-T90 of HLRZ J\"ulich.  J.J.
  and H.P. thank HLRZ J\"ulich for hospitality. The work was supported by DFG.
\end{ack}

\bibliographystyle{wunsnot}   


\end{document}

%% file: tab/statistics.tex
%
%
%
\begin{table}[tbp]
\hfuzz=1pt
\caption{
Numbers of measurements in the confinement phase in multiples of 1000. All
runs in which the two state signal was present are omitted from the
anylsis. They are denoted by a \dag\ in this table. The phase transition is at
$\beta_c=1.4223(2)$.}  
\begin{center}
\leavevmode
\footnotesize
\begin{tabular}{lrrrrr}
\hline
$\beta$ & $16^332$ & $20^340$ & $24^348$ & $28^356$ & $32^364$ \\
\hline
1.4     & 4.5      &          &          &          &          \\
1.405   & 5.0      &          &          &          &          \\
1.41    & 3.6      &          &          &          &          \\
1.412   & 3.0      &          &          &          &          \\
1.414   & 3.6      &          &          &          &          \\
1.416   & 3.0      &          &          &          &          \\
1.418   & 4.1      & 2.5      &          &          &          \\
1.419   & \dag     & 5.5      & 1.9      &          &          \\
1.42    & \dag     & 5.7      & 2.7      &          &          \\
1.4205  & \dag     & \dag     & 4.1      & 1.9      &          \\
1.4207  &          & \dag     & \dag     & \dag     & 1.1      \\
1.4210  &          &          &          & \dag     & \dag     \\
\hline
\end{tabular}
\end{center}
\label{tab:stat}
\end{table}

%
%

%% file: tab/symmetries.tex
%
%
%

\begin{table}
\caption{
  GB states observed in various channels with symmetry $R^{PC}$ and momentum
  $p=0,1$ (in units of $2\pi/L_s$) and their scaling behaviour. If two momenta
  are listed in one line, the masses have been assumed to be the same. 
  The ordering of the lines is the same as used in \cite{CoFr97b}.
  The reliability of the energy determination is indicated as ``quality''. 
  In the columns ``$\nu$'' and ``mass'' the scaling exponent of the GB mass
  and its approximate proportionality to one of the mass scales is
  indicated. The ratio of each mass to the corresponding mass scale is given
  in the column ``ratio''. The ratios at $\gamma=-0.2$ from \cite{CoFr97b} are
  shown for comparison. Since we used only GB masses $m_j(\beta)\leq 1.0$ in
  lattice units in the determination of the scaling behaviour, the amplitudes
  and therefore the ratios are not known for all of these channels.}
\begin{center}
\begin{tabular}{l|c|c|c|c|c}
\hline
$R^{PC}(p)$       & Quality  &$\nu$           & Mass                & Ratio  & Ratio    \\
&&&& $\gamma=-0.5$ & $\gamma=-0.2$ \\
\hline      
$A_1^{++}(0,1)$   &          & $\nu_{\rm g}$  & $m_{\rm g}$         & 1      & 1        \\       
$T_1^{-+}(1)$     &          &                &                     & 1.0(3) & 1.01(6)  \\
\cline{1-1}\cline{3-6}      
$A_2^{+-}(0)$     &          &                &                     & 1.0(1) & 1.00(4)  \\ 
$E^{+-}(0)$       &$+++$     &                &                     & 1.0(1) & 0.98(4)  \\ 
$T_1^{+-}(0,1)$   &          &                & $m_{\rm ng}$        & 1      & 1        \\ 
$T_2^{+-}(0,1)$   &          &                &                     & 1.0(1) & 1.00(4)  \\ 
$T_2^{--}(1)$     &          &                &                     &        & 1.04(5)  \\ 
\cline{1-2}\cline{4-6}      
$E^{++}(0,1)$     &$++$      &                & $\simeq2m_{\rm ng}$ & 2.2(2) & 2.06(9)  \\ 
$T_2^{++}(0,1)$   &          &                &                     & 2.1(2) & 2.07(9)  \\
\cline{1-2}\cline{4-6}      
$A_1^{-+}(0)$     &$?$       &                &                     &        & 2.88(13) \\ 
$A_1^{--}(0)$     &$?$       & $\nu_{\rm ng}$ &                     &        & 2.73(14) \\ 
$A_2^{++}(0)$     &$+$       &                &                     &        & 3.22(16) \\ 
$A_2^{--}(0)$     &$+$       &                &                     & 3.0(3) & 2.73(15) \\ 
$E^{--}(0)$       &$+$       &                &                     &        & 2.95(14) \\ 
$T_1^{++}(0,1)$   &$+$       &                &                     & 3.5(4) & 3.61(18) \\ 
$T_1^{--}(0,1)$   &$+$       &                & $\simeq3m_{\rm ng}$ & 3.0(3) & 2.94(13) \\ 
$T_2^{-+}(0)$     &$+$       &                &                     &        & 3.10(14) \\ 
$T_2^{-+}(1)$     &$+$       &                &                     &        & 2.63(12) \\ 
$T_2^{--}(0)$     &$+$       &                &                     &        & 3.25(16) \\ 
$A_1^{+-}(0)$     &$+$       &                &                     & 3.0(3) &          \\
$A_2^{-+}(0)$     &$+$       &                &                     &        &          \\
$E^{-+}(0)$       &$+$       &                &                     & 2.8(3) &          \\
$T_1^{-+}(0)$     &$+$       &                &                     & 3.8(4) &          \\
\hline      
\end{tabular}
\end{center}
\label{tab:channels}
\end{table}

%
%